\documentclass[letterpaper, conference]{ieeeconf}
\IEEEoverridecommandlockouts
\usepackage{balance}
\usepackage{cite}
\usepackage{float}
\usepackage{multicol,multirow}
\usepackage{makecell,booktabs}
\usepackage{hyperref}
\hypersetup{    
    colorlinks=true,
    linkcolor=black,
    anchorcolor=black,
    citecolor=black,
    filecolor=black,
    menucolor=black,
    runcolor=black,
    urlcolor=black,
}

\usepackage{subcaption} % for subplot captions
\usepackage[utf8]{inputenc}

\usepackage{graphicx}
\usepackage{amsfonts}
\usepackage{amssymb}
\usepackage{amsmath,mathtools}
\usepackage{array}
\newcolumntype{P}[1]{>{\centering\arraybackslash}p{#1}}
\newcolumntype{L}[1]{>{\raggedright\arraybackslash}p{#1}}
\newcolumntype{R}[1]{>{\raggedleft\arraybackslash}p{#1}}
\usepackage{color}

\usepackage[labelsep=period]{caption}
\usepackage{bbm}
\usepackage{bm}
\usepackage{comment}
\usepackage{forest}
\usepackage{dblfloatfix}

\usepackage{etoolbox}

\title{\bf 
    Identification of Intraday False Data Injection Attack on\\DER Dispatch Signals
}
\author{
    Jip Kim, Siddharth Bhela, James Anderson, and Gil Zussman
    \thanks{Jip Kim, James Anderson, and Gil Zussman are with the Department of Electrical Engineering, Columbia University, New York, NY 10027, USA. Emails: {\tt\small \{jk4564, ja3451, gz2136\}@columbia.edu}. 
    Siddharth Bhela is with Siemens Technology, Princeton, NJ 08540, USA. Email: {\tt\small siddharth.bhela@siemens.com}.
    This material is based upon work supported by the U.S. Department of Energy's Office of Energy Efficiency and Renewable Energy (EERE) under the Solar Energy Technology Office Award Number DE-EE0008769. 
    The views expressed herein do not necessarily represent the views of the U.S. Department of Energy or the United States Government.
    }    
}

\begin{document}

\maketitle

\begin{abstract}
    The urgent need for the decarbonization of power girds has accelerated the integration of renewable energy. Concurrently the increasing distributed energy resources (DER) and advanced metering infrastructures (AMI) have transformed the power grids into a more sophisticated cyber-physical system with numerous communication devices. While these transitions provide economic and environmental value, they also impose increased risk of cyber attacks and operational challenges.  This paper investigates the vulnerability of the power grids with high renewable penetration against an intraday false data injection (FDI) attack on DER dispatch signals and proposes a kernel support vector regression (SVR) based detection model as a countermeasure. The intraday FDI attack scenario and the detection model are demonstrated in a numerical experiment using the HCE 187-bus test system.
\end{abstract}

\section{Introduction}\label{Sec:Intro}
\subsection{Background}
\begin{figure}[t]
    \centering    
    \vspace{-0mm}
    \includegraphics[width=\columnwidth]{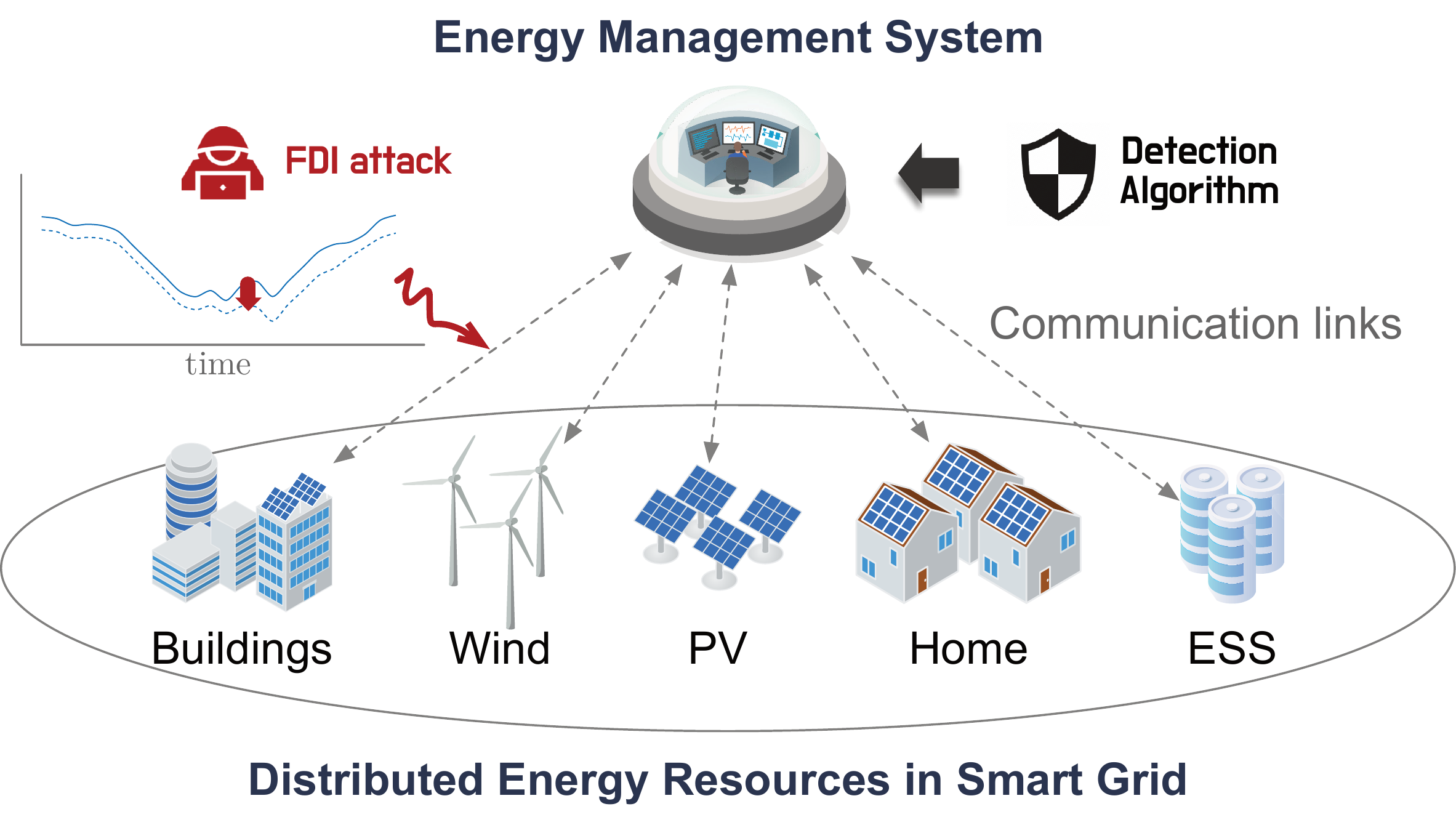}
    \vspace{-6mm}
    \caption{\small 
    Various DERs and connecting communication links to the EMS. Potential intraday FDI attacks targeting the communication links and detection algorithm for EMS are also illustrated.
    \vspace{-5mm}}
    \label{fig:intro}
\end{figure}

Along with the rapid deployment of distributed energy resources (DERs), power system operation heavily relies on information communication technologies (ICT) as shown in Fig.~\ref{fig:intro}~\cite{pong2021cyber}. DERs such as energy storage and small-scale generators receive dispatch signals from the energy management system (EMS) and  consumer-side resources such as electric vehicles and demand response also contribute to the increase in the number of ICT devices involved in power grid operations. While the DERs generally provide economic benefits to the electricity suppliers and consumers, the power grid is more exposed to cyber attacks through the ICT devices. For example, in 2015, a coordinated cyber attack against the ICT network and devices in the distribution grid caused blackouts in three different regions in Ukraine for more than seven hours \cite{liang20162015}.

Meanwhile, renewable integration and decarbonization of the power grid have significantly increased the system's operational difficulties. The intermittent nature of clean, but weather-dependent energy resources such as wind turbine and photovoltaic (PV) generators increase the need in ramping resources to handle the variability and system reserves to address the subsequent uncertainty. At the same time, the profitability of  conventional flexibility resources such as coal and gas power plants has been aggravated, making it more challenging to secure an adequate amount of balancing resources \cite{Udive_coal}. In practice, California ISO which is known for its aggressive PV integration is facing the so-called \emph{duck curve} phenomenon which refers to the shape of the net load (total demand minus renewable generation) profile with a high solar penetration including a rapid ramp-up due to the reduction in solar generation around sunset. The maximum three-hour ramping requirement was predicted to be 13GW/3hr for the year 2020 in 2013, but the actual value turned out to be 17GW/3hr \cite{CAISO_today}. This trend is currently observed mostly among the renewable pioneers such as power grids in California and Texas, and will accelerate along with the installation of more PV generation. The steep net load ramp-up requires the power grid to operate with small system reliability and resilience margins, and in turn, increases the system vulnerability to unexpected events such as cyber attacks.

\subsection{Related work}
A range of cyber attacks in power grids have been investigated in the literature \cite{ten2008vulnerability,liu2019resilience,chen2019distributed,jiang2013spoofing,pan2015developing,huang2018online,soltan2018react,aoufi2020survey,zhang2021smart,rahman2013false,isozaki2015detection,soltan2016power}. A \emph{denial of service} (DoS) attack that renders the entire communication network in the grid unavailable by injecting meaningless packets was presented in \cite{ten2008vulnerability,liu2019resilience,chen2019distributed}. A \emph{replay attack} which records the reading of sensors and repeats these historical values was studied in \cite{jiang2013spoofing,pan2015developing,huang2018online,soltan2018react}. The most extensively examined type of attack is a \emph{false data injection} (FDI) attack \cite{rahman2013false,isozaki2015detection,khanna2017bi}. FDI attacks attempt to change the system state by injecting falsified data into the communication devices. The aforementioned example in Ukraine falls into this category. Interested readers are referred to review papers \cite{aoufi2020survey,zhang2021smart} for general cyber attacks in power grids. We narrow our attention to FDI attacks in this paper.

FDI attacks (in power grids) can be classified by the target functionality. Rahman \textit{et al.} \cite{rahman2013false} studied FDI attacks against the state estimation module which can be critical as incorrect situational awareness causes a malfunction in reliability applications such as contingency analysis. Isozaki \textit{et al.} \cite{isozaki2015detection} inspected FDI attacks on voltage regulation control causing irregular tap changing by manipulating a limited number of load measurements and Choeum \textit{et al.} \cite{choeum2019oltc} demonstrated a similar FDI attacks on Volt/VAR control which can deteriorate the power quality. Also, Khanna \textit{et al.} \cite{khanna2017bi} investigated FDI attacks on optimal power flow by falsifying load measurements so that the resulting generation dispatch is not N-1 security compliant. However, to the best of our knowledge, there is no existing work that analyzes the FDI attacks associated with the DER dispatch signals.

As diverse as the type of FDI attacks are, proposed detection algorithms are equally diverse \cite{pasqualetti2013attack,musleh2019survey}. Detection methods can be broadly divided into model-based and model-free, where model-based methods  use power system state estimation, exploiting  network information and physical system knowledge (e.g., grid topology, line impedances, electricity demand, etc.) to detect anomalies in the observation \cite{ashok2016online,teixeira2010cyber}. In contrast, model-free approaches  exploit recent advances in machine learning techniques such as classification and clustering \cite{cui2020detecting}. There are, however, limited detection approaches capable of capturing temporal characteristics. A short-term state forecasting model in \cite{zhao2015short} takes temporal correlation among different nodal states into account to detect FDI attacks and Karimipour \textit{et al.} \cite{karimipour2017robust} presented a dynamic state estimation method accounting for multiple time frames using Kalman filtering. It is noteworthy that none of the existing work has focused on the counter measurements and system vulnerability assessment for intraday FDI attacks.

Support vector regression (SVR) has gained popularity for time series forecasting in different domains. Thissen \textit{et al.} \cite{thissen2003using} suggested that the SVR can model nonlinear relations and generate time series predictions. Salcedo-Sanz \textit{et al.} \cite{salcedo2011short} showed how SVR can be used for wind speed prediction, and He \textit{et al.} \cite{he2008model} applied SVR to electricity load prediction. In addition, Feng \textit{et al.} \cite{feng2018adaptive} utilized a multiple kernel SVR to capture both local and global information and applied to the traffic flow prediction. Building on these applications, we aim to exploit SVR for considering multi-temporal correlation in the system status and dispatch signals and identifying  time-series FDI attacks on power systems.

\subsection{Contribution}
The  contributions of this paper are two-fold. First, this paper reveals the vulnerability of the low-carbon power systems against the intraday FDI attacks by developing an attacker model that is composed of two optimization models: dispatch prediction model and dispatch falsification model. Based on the historical values and the collected knowledge of the target power grid, the dispatch prediction model mimics the functionality of the EMS and predicts the dispatch signals for DERs. Once the dispatch signal prediction is made, the dispatch falsification model solves an optimization problem to determine how to falsify the dispatch signals between DERs and EMS so that the accumulated deviations in DER outputs exceed the system security margin and cause the system power shortage. Second, this paper proposes a kernel SVR based detection model to enhance the reliability of the power grid against cyber attacks. The kernel SVR takes the input of monitored data comprised of multi-interval dispatch signals and the corresponding network status (nodal voltage magnitudes and phase angles) and predicts the system margin of the time of interest (e.g., two hours ahead). If the predicted margin drops below the threshold, the detection model notifies the system operator. To increase the performance of the kernel SVR, the training data is normalized by feature and transformed into a feature space using a kernel trick.

\begin{figure}[t]
    \centering    
    \vspace{-0mm}
    \includegraphics[width=0.7\columnwidth]{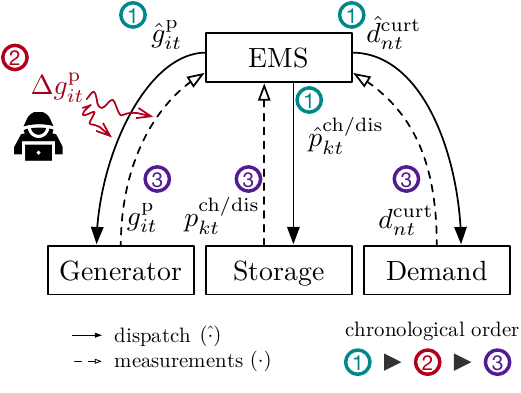}
    \vspace{-3mm}
    \caption{\small 
    Dispatch (solid lines) and measurement (dashed lines) signals between EMS and energy resources. The intervention of the attacker is marked in red color.
    \vspace{-0mm}}
    \label{fig:FDIA_g}
    \centering    
    \vspace{-0mm}
    \includegraphics[width=\columnwidth]{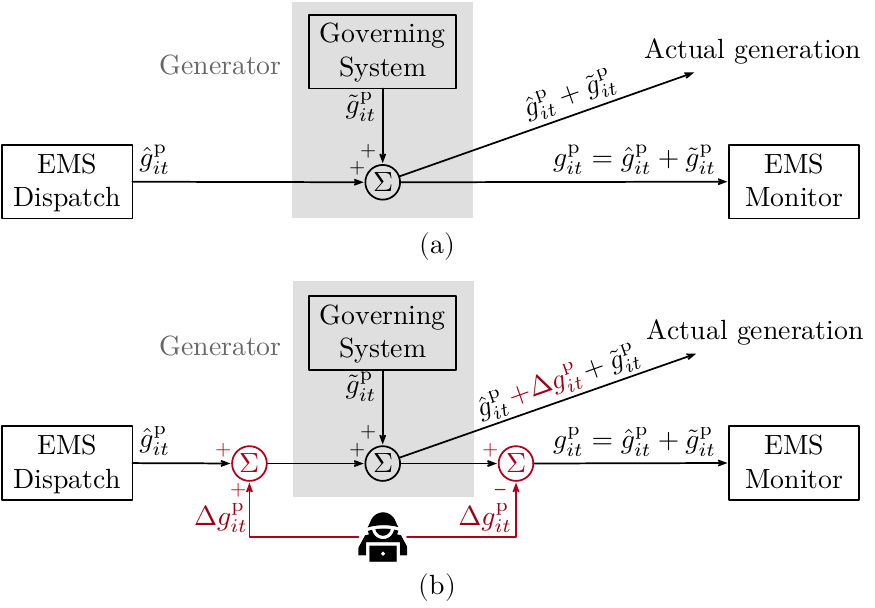}
    \vspace{-7mm}
    \caption{\small
        Dispatch signal flow of (a) normal operation  and (b) operation under intraday FDI attack on generation dispatch.
        \vspace{-4mm}}
    \label{fig:signal_flow}
\end{figure} 

\section{Intraday FDI attack}\label{sec:FDIA}
A key contribution of this work is to highlight the vulnerability of power systems to intraday FDI attacks. This section describes a gradually pervasive FDI attack that can be carried out during the evening net demand ramping-up periods. The attacker manipulates the dispatch signals from the EMS so that the power output of the various DERs (e.g., distributed generator, energy storage, demand response) deviates from the desired output. Once the accumulated deviation in the power output is larger than the system flexibility (reserve and ramping resources), then the supply cannot follow the demand ramp-up which can cause  system-wide power outages and potentially cascading failures. This attack can be critical as the net demand continues to increase rapidly over the hours which makes the recovery process from the outage more challenging.

\subsection{Overview of intraday FDI attack}
The model in the following subsections is generic to intraday FDI attacks against any type of DERs that receive dispatch signals from EMS, while Figs.~\ref{fig:FDIA_g}\textendash\ref{fig:signal_flow} use the attack on generation dispatch for illustrative purposes.
Figure~\ref{fig:FDIA_g} illustrates the intraday FDI attack against the interactions between the EMS and the energy resources. First the attacker collects dispatch data and produces the falsification signal ($\Delta g^{\mathrm{p}}_{it}$) for the target time window. The actual attack is carried out by falsifying the dispatch going into the energy resource ($\hat{g}^{\mathrm{p}}_{it}$) and the monitoring signal ($g^{\mathrm{p}}_{it}$) into the EMS simultaneously. By doing so, the monitoring signal will appear the same from the EMS' perspective as $\hat{g}^{\mathrm{p}}_{it}+ \tilde{g}^{\mathrm{p}}_{it}$ under the normal operation in Fig.~\ref{fig:signal_flow}(a) and the one under the intraday FDI attack in Fig.~\ref{fig:signal_flow}(b), while the actual generation is perturbed by $\Delta{g}^{\mathrm{p}}_{it}$ from the original dispatch regardless of presence of the governor adjustment. Other resources that receive dispatch signals from EMS such as energy storage, demand response can be exposed to the same type of intraday FDI attack.

\subsection{Dispatch prediction model}\label{subsec:prediction}
To generate multi-interval falsifying dispatch signals, the attacker needs to predict the original dispatch signals from the EMS. This requires knowledge of the power grids such as network topology and nodal electricity demands. The attack model in this paper is assumed to have perfect knowledge of necessary information (network topology, line impedances, and thermal limis -- see Table~\ref{tab:assumption}), which can be considered as the most precise attack scenario.\footnote{In the 2015 Ukraine case \cite{liang20162015} the network information was obtained by reconnaissance operations.} The type of information needed for the attack model is summarized in Table~\ref{tab:assumption}.\footnote{While the reserve allocation for each balancing unit is marked as unknown, the attack model requires the knowledge of the total reserve.} The dispatch prediction model mimics the power grid operation and is formulated as the second-order cone problem:
\begingroup
\allowdisplaybreaks
\begin{subequations}\label{Eq:AttackerPred}
\begin{align}
    \begin{split}
    & 
    \{
        \hat{g}^{\mathrm{p}}_{it}, 
        \hat{d}^{\mathrm{p,curt}}_{nt},
        \hat{p}^{\mathrm{ch}}_{kt},
        \hat{p}^{\mathrm{dis}}_{kt},
        \hat{r}_{it} 
    \}\in \arg\\
    &
    \!\!\min_{\Xi^{\mathrm{P}}} 
        \sum_{t\in{\cal T}}
        \sum_{i\in{\cal I}}
            c^{\mathrm{g}}_i(\hat{g}^{\mathrm{p}}_{it}) +
            c^{\mathrm{d}}_n(\hat{d}^{\mathrm{p,curt}}_{nt}) +
            c^{\mathrm{r}}_i(\hat{r}_{it})
    \end{split}\label{Eq:pred_obj}\\
    & \mathrm{subject~to:}\nonumber\\
    \begin{split}
        & f_{l \vert o(l) = n,t}^{\mathrm{p}} 
            \!- \hspace{-2.0mm} \sum_{l \vert r(l) = n } \hspace{-2.0mm} 
                (f^{\mathrm{p}}_{lt} \!-\! a_{lt} R_l) 
            \!-\!\sum_{i\in{\cal I}_n} 
                \hat{g}^{\mathrm{p}}_{it} 
            + D^{\mathrm{p}}_{nt}  
            \\& \hspace{9mm}
            - \hat{d}^{\mathrm{p,curt}}_{nt}
            + G_n u_{nt}  
            = 0,
                \quad \forall n \in {\cal N},~ t \in {\cal T},
    \end{split}
        \label{Eq:pred_P_balance}\\
    \begin{split}
        & f_{l \vert o(l) = n,t}^{\mathrm{q}} 
            \!- \hspace{-2.0mm} \sum_{l \vert r(l) = n } \hspace{-2.0mm}
                (f^{\mathrm{q}}_{lt} \!-\! a_{lt} X_l)
            \!-\!\sum_{i\in{\cal I}_n}
                \hat{g}^{\mathrm{q}}_{it}
            + D^{\mathrm{q}}_{nt} 
            \\& \hspace{9mm}
            - \Gamma_n \hat{d}^{\mathrm{p,curt}}_{nt}
            - B_n u_{nt} 
            = 0,
                \quad\forall n \in {\cal N},~ t \in {\cal T},
    \end{split}
        \label{Eq:pred_Q_balance}\\      
    \begin{split}
        & u_{o(l),t}\!
            -\!2\!\left(R_l f_{lt}^{\mathrm{p}} 
            \!+\! X_l f^{\mathrm{q}}_{lt} \right)
            + a_{lt} (R_l^2 + X_l^2)
                =\! u_{r(l),t},
            \\& \hspace{9mm}
                \forall l \in {\cal L}, ~ t \in {\cal T},
    \end{split}\label{Eq:pred_DistFlow1}\\
    & 
    \begin{bmatrix}  
        a_{lt} + u_{o(l),t}\\ 
        a_{lt} - u_{o(l),t}\\ 
        2f^{\mathrm{p}}_{lt}\\ 
        2f^{\mathrm{q}}_{lt} 
    \end{bmatrix} \in {\cal K}^4, 
        \quad \forall l \in {\cal L}, ~ t \in {\cal T},
        \label{Eq:pred_SOC}\\
    & 
        \begin{bmatrix} 
            F_l \\
            \hat{f}^{\mathrm{p}}_{lt}\\
            \hat{f}^{\mathrm{q}}_{lt}
        \end{bmatrix},~
        \begin{bmatrix}
            F_l \\ 
            \hat{f}^{\mathrm{p}}_{lt} - \hat{a}_{lt} R_l\\
            \hat{f}^{\mathrm{q}}_{lt} - \hat{a}_{lt} X_l
        \end{bmatrix}
            \in {\cal K}^3,
            \quad  \forall l \in {\cal L}, ~ t \in {\cal T}
            \label{Eq:pred_LineLimit}\\
    & \sum_{i \in {\cal I}}
        \hat{r}_{it} 
        \ge
            K^{\mathrm{r,sys}} \sum_{n \in {\cal N}} D^{\mathrm{p}}_{nt}
                ,\quad \forall t \in {\cal T},
                    \label{Eq:pred_reserve_req}\\
    & R^{\mathrm{dn}}_i 
        \le
            \hat{g}^{\mathrm{p}}_{it} - \hat{g}^{\mathrm{p}}_{i,t-1}
        \le
            R^{\mathrm{up}}_i 
                ,\quad \forall i \in {\cal I}^{\mathrm{C}}, t \in {\cal T},
                    \label{Eq:pred_ramp_req}\\
    & 
        \underline{G}^{\mathrm{p}}_i
            \le \hat{g}^{\mathrm{p}}_{it} + \hat{r}_{it}
            \le \overline{G}^{\mathrm{p}}_i,
                \quad \forall i \in {\cal I}, t \in {\cal T},
                    \label{Eq:pred_PGbound}\\       
    & 
        \underline{G}_i^{\mathrm{q}} 
            \le \hat{g}^{\mathrm{q}}_{it}
            \le \overline{G}^{\mathrm{q}}_i,
                \quad \forall i \in {\cal I}, t \in {\cal T},
                    \label{Eq:pred_QGbound}\\
    & 
        \underline{U}_n 
            \le \hat{u}_{nt} 
            \le \overline{U}_n, 
                \quad \forall n \in {\cal N}, t \in {\cal T},
                    \label{Eq:pred_Vbound}\\
    & \underline{D}^{\mathrm{p,curt}}_{nt}
        \le
            \hat{d}^{\mathrm{p,curt}}_{nt}
        \le
            \overline{D}^{\mathrm{p,curt}}_{nt}
                ,\quad\forall n \in {\cal N},
                    \label{Eq:pred_dcurt_bounds}\\
    & \hat{e}_{kt}
        = \hat{e}_{k,t-1} 
            + 
                \hat{p}_{kt}^{\mathrm{ch}}  \aleph^{\mathrm{ch}}
                - \hat{p}_{kt}^{\mathrm{dis}}  / \aleph^{\mathrm{dis}} 
            ,
        \quad\forall k \in {\cal K}, t \in {\cal T},\hspace{-2mm}
    \label{Eq:pred_ESS_dynamic}\\
    & \underline{E}_{kt} \le \hat{e}_{kt} \le \overline{E}_{kt}, 
            \quad\forall k \in {\cal K}, t \in {\cal T},
        \label{Eq:pred_ESS_capacity_limit}\\
    & \underline{P}_k \le \hat{p}_{kt}^{\mathrm{ch}} \cdot \aleph^{\mathrm{ch}} \le \overline{P}_k,
        \quad\forall k \in {\cal K}, t \in {\cal T},
    \label{Eq:pred_pch_limit}\\
    & \underline{P}_k \le \hat{p}_{kt}^{\mathrm{dis}} / \aleph^{\mathrm{dis}} \le \overline{P}_k ,
        \quad\forall k \in {\cal K}, t \in {\cal T},
    \label{Eq:pred_pdis_limit}
\end{align}
\end{subequations}
\endgroup
where $\Xi^{\mathrm{P}} \coloneqq \{ 
        \hat{g}^{\mathrm{p}}_{it}, 
        \hat{g}^{\mathrm{q}}_{it}, 
        \hat{r}_{it},
        \hat{d}^{\mathrm{p,curt}}_{nt},
        \hat{e}_{kt},
        \hat{p}^{\mathrm{ch}}_{kt},
        \hat{p}^{\mathrm{dis}}_{kt},
        \hat{a}_{lt}, 
        \hat{u}_{nt} 
            \ge 0, \\
        \hat{d}^{\mathrm{p,net}}_{nt},
        \hat{f}^{\mathrm{p}}_{lt}, 
        \hat{f}^{\mathrm{q}}_{lt} 
                \in \mathbb{R}\}$
and the second-order-cone is defined as $
    {\cal K}^n \coloneqq 
        \{x \in {\mathbb{R}}^{n} \vert x_1 \ge {\sqrt{x_2^{2} + \cdots + x_n^{2}}} \}
    $. 
Power grid is defined with lines $l\in{\cal L}$ and nodes $n\in{\cal N}$ while time set is denoted as $t\in{\cal T}$. The objective function in \eqref{Eq:pred_obj} minimizes the total cost of generation, load curtailment and reserve. Equations \eqref{Eq:pred_P_balance}\textendash\eqref{Eq:pred_SOC} are second-order-conic relaxation of the AC power flow equations \cite{farivar2013branch}, where $o(l)$ and $r(l)$ denote the sending and receiving buses of line $l$. Forward and backward line flow limits are enforced in \eqref{Eq:pred_LineLimit} with line capacity $F_l$. Given reserve requirement parameter $K^{\mathrm{r,sys}}$, the minimum system reserve is set proportional to the total demand in \eqref{Eq:pred_reserve_req}. Ramping constraints of flexible generation units ($i\in{\cal I}^{\mathrm{C}}$) are imposed in \eqref{Eq:pred_ramp_req}. The active and reactive power limits of generators are enforced in \eqref{Eq:pred_PGbound} and \eqref{Eq:pred_QGbound}. Equations \eqref{Eq:pred_Vbound} and \eqref{Eq:pred_dcurt_bounds} limit lower and upper bounds of nodal voltage magnitudes and load curtailments respectively. Energy storage operation is modeled in \eqref{Eq:pred_ESS_dynamic}\textendash\eqref{Eq:pred_pdis_limit} where the charging and discharging decisions are denoted as $\hat{p}^{\mathrm{ch}}_{kt}$ and $\hat{p}^{\mathrm{dis}}_{kt}$. The inter-temporal relationship of the state of charge $\hat{e}_{kt}$ is defined in \eqref{Eq:pred_ESS_dynamic} with charging and discharging efficiency parameters, $\aleph^{\mathrm{ch}}$ and $\aleph^{\mathrm{dis}}$. The lower and upper bound constraints for the charging and discharging power are in \eqref{Eq:pred_pch_limit} and \eqref{Eq:pred_pdis_limit}.

\subsection{Dispatch falsification model}\label{subsec:falsification}
Once the prediction of the dispatch is made, the attacker can generate falsification signals. The falsification targets are constrained by the number and type of access points. For simplicity, it is assumed that the FDI attack would be carried out using only a single type of DERs (e.g., generator, storage or demand in Fig.~\ref{fig:FDIA_g}). Then the dispatch falsification model is formulated with generic dispatch notation $x_{kt}$ for unit $k$ and attack time $t\in\mathcal T^{a}$ as follows:
\begingroup
\allowdisplaybreaks
\begin{subequations}\label{Eq:AttackerModel}
\begin{align}
    & 
    \min_{\Delta x_{kt}} 
        \sum_{t\in{\cal T}} \sum_{k\in{\cal K}}
        \left(
           \lvert \Delta x_{kt} \rvert^2
        + \rho 
           \lvert \Delta x_{kt} \hspace{-1mm}- \Delta x_{k,t-1} \rvert^2
        \right)\label{Eq:Attacker_Obj}\\
    & \mathrm{subject~to:}\nonumber\\
    & 
        \lvert 
        \sum_{k \in {\cal K}}
                \Delta \hat{x}_{kt}
                \rvert
        \ge 
            K^{\mathrm{a}}_t \sum_{i \in {\cal I}}
                \hat{r}_{it},
                \quad \forall t \in {\cal T}^{\mathrm{a}}
                    \label{Eq:Attacker_res}\\
    & 
        \lvert
            \sum_{t\in{\cal T}^{\mathrm{a}}}\sum_{k\in{\cal K}}
            \Delta x_{kt}
        \rvert
        \ge
            \sum_{t\in{\cal T}^{\mathrm{a}}} \sum_{i\in{\cal I}}
                \hat{r}_{it},
                    \label{Eq:Attacker_ramp}\\
    & - \epsilon^{\mathrm{a}} \hat{x}_{kt}
        \le \Delta x_{kt}
        \le \epsilon^{\mathrm{a}} \hat{x}_{kt},
            \quad \forall k \in {\cal K}, t \in {\cal T}^{\mathrm{a}},
                \label{Eq:Attacker_DelXbounds}\\
    & - \underline{X}_{kt}
        \le \hat{x}_{kt} + \Delta x_{kt}
        \le \overline{X}_{kt},
            \quad \forall k \in {\cal K}, t \in {\cal T}^{\mathrm{a}},
                \label{Eq:Attacker_Xbounds}
\end{align}
\end{subequations}
\endgroup
Given the prediction of the target dispatch (e.g., $\hat{x}_{kt} \coloneqq \hat{g}^{\mathrm{p}}_{it}$ for generator, $\hat{x}_{kt} \coloneqq \hat{p}^{\mathrm{ch/dis}}_{kt}$ for storage, and $\hat{x}_{kt} \coloneqq \hat{d}^{\mathrm{p,curt}}_{nt}$ for load curtailment), the attacker determines the falsification signal (e.g., $\Delta x_{kt} = \Delta g^{\mathrm{p}}_{it}$). The objective function in \eqref{Eq:Attacker_Obj} minimizes the sum of the squared size of the attack and the temporal smoothness regularization term\footnote{The regularization term in \eqref{Eq:Attacker_Obj} can be extended to account for other dimensions such as geographical locations (i.e., similar deviations in nearby dispatch signals).} with penalty parameter $\rho$. Constraint \eqref{Eq:Attacker_res} sets the impact of the attack achieves the target deviation with user-defined parameter $K^{\mathrm{a}}_t$ for each time interval and \eqref{Eq:Attacker_ramp} ensures the accumulated deviation in the dispatch during the attacking windows $t\in{\cal T}^{\mathrm{a}}$ exceeds the system reserve. The lower and upper bounds for individual falsification signals are given in \eqref{Eq:Attacker_DelXbounds} and \eqref{Eq:Attacker_Xbounds} where $\epsilon^{\mathrm{a}}$ is an user-defined parameter to confine the attack size and the original dispatch bounds are $\underline{X}_{kt}$ and $\overline{X}_{kt}$.

\begin{table}[tp]
\centering
\captionsetup{justification=centering, labelsep=period, font=footnotesize, textfont=sc}
\caption{Assumptions for the Attack and Detection Models}
    \vspace{-2mm}
    \begin{center}
    \begin{tabular}{L{35mm} | P{20mm} | P{20mm} }
        \toprule
        \centering              
        Name& Attack Model & Detection Model \\
        \midrule
        Network information:&&\\
        \hspace{4mm} topology (${\cal N}, {\cal L}$) & \checkmark & \checkmark\\
        \hspace{4mm} line impedance ($R_l,X_l$) & \checkmark & \checkmark\\
        \hspace{4mm} line thermal limit ($F_l$) & \checkmark & \checkmark\\
        \midrule
        Dispatch signals:&&\\
        \hspace{4mm} generation output (${g}^{\mathrm{p}}_{it}$) & \textendash & P~\&~A\\
        \hspace{4mm} load curtailment (${d}^{\mathrm{curt}}_{nt}$) & \textendash & P~\&~A\\
        \hspace{4mm} storage dispatch (${p}^{\mathrm{ch/dis}}_{kt}$) & \textendash & P~\&~A\\
        \hspace{4mm} reserve (${r}_{it}$) & \textendash & P~\&~A\\
        \midrule
        Measurements:&&\\
        \hspace{4mm} nodal demand (${D}^{\mathrm{p}}_{nt}$) & P & P~\&~A\\
        \hspace{4mm} nodal voltage ($v_{nt},\theta_{nt}$) & \textendash &  \checkmark\\
        \bottomrule
        \multicolumn{3}{l}{\color{black}$^*$ P:~prediction,~A:~actual,~\checkmark:~ assumed to be known,~\textendash:~unknown}
    \end{tabular}
    \end{center}
    \label{tab:assumption}
    \vspace{-8mm}
\end{table}

\section{FDI attack detection with kernel SVR}\label{sec:SVR}
This section reviews the kernel SVR algorithm and describes how it can be used for the intraday FDI attack detection. In order to detect the intraday FDI attack carried out over several hours, the detection model should be able to capture the temporal changes in dispatch signals as well as the current values. To do so, the detection model requires access to the dispatch signals and network status for the monitoring windows ${\cal T}^{\mathrm{m}}$ ({in the sequel we shall use a 6-hour window}) and the system margin at the time of interest $T^{\mathrm{pred}}$ ({in our examples, the time of interest is set to 2 hours after the monitoring window closes}) as summarized in Table~\ref{tab:assumption}. Note that this data can be provided by the EMS.

\subsection{Kernel Support Vector Regression}
Kernel SVR is a generalization of Kernel support vector machine for real-value function estimation, commonly equipped with $\epsilon$-insensitive loss function ({sometimes referred to as soft-margin loss function}) \cite{smola2004tutorial}. A kernel {function maps} the original input data into feature space (i.e., $x^i \mapsto \phi(x^i)$) {through the use of inner products}, which is compatible for the SVR model fitting. The common selection of a kernel function is Gaussian radial basis kernel function (RBF) in \eqref{eq:kernel_rbf} and polynomial kernel function of degree $d$ in \eqref{eq:kernel_poly} (See \cite{rohmah2021comparison} for more details on the kernel selection for SVR):
\begin{subequations}\label{Eq:kernel}
\begin{align}
    & k(x,y) \coloneqq e^{-\frac{\lVert x- y\rVert^2}{2\sigma^2}}\label{eq:kernel_rbf}\\
    & k(x,y) \coloneqq (x^\top y)^d\label{eq:kernel_poly}
\end{align}
\end{subequations}

Given input data $x^i$ with user-defined parameter $C$ and feature map $\phi(\cdot)$, the Kernel SVR for predicting output: $y_i\in \mathbb{R}$ can be modeled as in \eqref{Eq:SVR_original}:
\begin{subequations}\label{Eq:SVR_original}
\begin{align}
    \min_{w,b,\xi^{+}_i, \xi^{-}_i}\quad &\lVert w \rVert + C \sum_{i=1}^{l} (\xi^{+}_i + \xi^{-}_i)\label{eq:svr_obj}\\
    \mbox{s.t.}\quad& (w^{\top} \phi(x^i) + b) - y_i \le \epsilon + \xi^+_i,\quad \forall i\label{eq:svr_con1}\\
    &y_i - (w^{\top} \phi(x^i) + b)  \le \epsilon + \xi^-_i,\quad \forall i\label{eq:svr_con2}\\
    &\xi^+_i, \xi^-_i \ge 0, \quad \forall i \label{eq:svr_con3}
\end{align}
\end{subequations}
The objective function in \eqref{eq:svr_obj} minimizes the sum of the norm of $\omega$ and the loss terms, where $\omega^\top x + b = 0$ is a decision boundary and the size of margin is $\frac{2}{\lvert\lvert \omega \rvert\rvert}$ (i.e., the margin is maximized). {Points within an $\epsilon$ distance of the support vector do not contribute to the cost.} {Although not pursued here, form a computational perspective, it is often favorable to work with the dual formulation of~\eqref{Eq:SVR_original} \cite{smola2004tutorial}.}

\subsection{Kernel SVR detection model}

\begin{figure}[t]
    \centering    
    \vspace{-0mm}
    \includegraphics[width=\columnwidth]{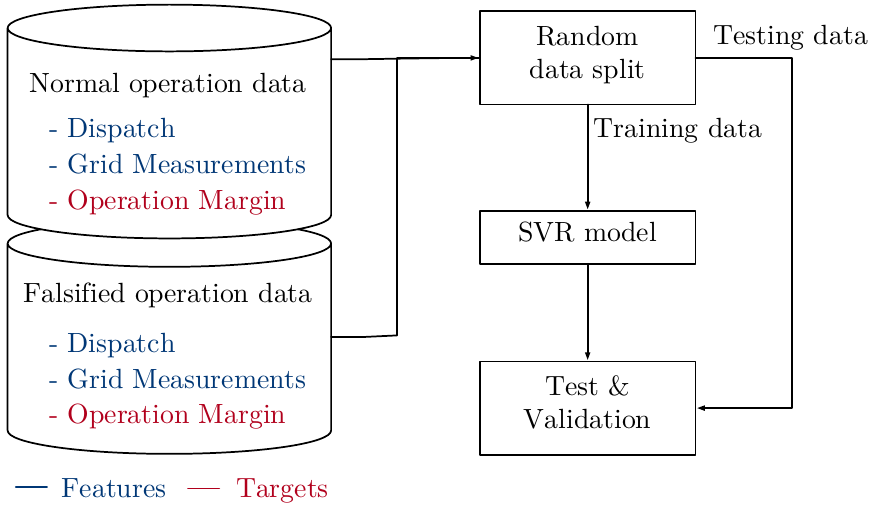}
    \vspace{-7mm}
    \caption{\small 
        Training and testing process of the kernel SVR detection model with the normal operation data and falsified dispatch data. 
        \vspace{-0mm}}
    \label{fig:data_gen}
\end{figure}

Figure~\ref{fig:data_gen} shows the process of training the kernel SVR for detecting the intraday FDI attacks. To train the SVR model in \eqref{Eq:SVR_original}, we first generate \textit{normal operation data} with supply and demand side uncertainties and \textit{falsified operation data}. The supply and demand uncertainties are modeled as a random variable drawn from a Gaussian distribution. In practice, historical data can be used for generating a probability distribution of the supply and demand uncertainties which will increase the performance of the kernel SVR. The power grid measurement data (voltage magnitudes and phase angles) are obtained by solving the power flow equations with fixed power injections based on the supply and demand data. Synthetic falsified operation data can be generated in the same way, while the dispatch signals are fixed as the falsified values from the dispatch falsification model in Section~\ref{subsec:falsification}.

Once the training data is prepared, the input vector $x^i$ for each observation $i$ is constructed which is comprised of prediction ($\hat{\cdot}$) and actual (${\cdot}$) values of energy resources over the user-defined monitoring time windows ($t\in{\cal T}^{\mathrm{m}}$) and the system status (voltage magnitudes and phase angles):
\begin{subequations}
\begin{align}
    \begin{split}
    &    x^i=\begin{bmatrix}
        [ \hat{g}^{\mathrm{p}}_{it}]_{\forall i\in {\cal I}, t \in {\cal T}^{\mathrm{m}}}\\
        [ {g}^{\mathrm{p}}_{it}]_{\forall i\in {\cal I}, t \in {\cal T}^{\mathrm{m}}}\\
        [ \hat{d}^{\mathrm{curt}}_{nt}]_{\forall n\in {\cal N}, t \in {\cal T}^{\mathrm{m}}}\\
        [ {d}^{\mathrm{curt}}_{nt}]_{\forall n\in {\cal N}, t \in {\cal T}^{\mathrm{m}}}\\
        [ \hat{p}^{\mathrm{ch/dis}}_{kt}]_{\forall k\in {\cal K}, t \in {\cal T}^{\mathrm{m}}}\\
        [ {p}^{\mathrm{ch/dis}}_{kt}]_{\forall k\in {\cal K}, t \in {\cal T}^{\mathrm{m}}}\\
        [ v_{nt}]_{\forall n\in {\cal N}^{\mathrm{s}}, t \in {\cal T}^{\mathrm{m}}}\\
        [ \theta_{nt}]_{\forall n\in {\cal N}^{\mathrm{s}}, t \in {\cal T}^{\mathrm{m}}}\\
    \end{bmatrix}
    \in {\mathbb{R}}^{ d },~ y_i \in \mathbb{R}\\
    &\text{where}~ d = (2 \lvert {\cal I} \rvert + 2\lvert {\cal N} \rvert +  2 \lvert {\cal K} \rvert + 2\lvert {\cal N}^{\mathrm{s}} \rvert) \lvert{\cal T}^{\mathrm{m}}\rvert
    \end{split}
\end{align}
Then the corresponding output ($y_i$) of the kernel SVR is set as the system operation margin defined as the minimum of the remaining up-ward and down-ward reserves at the time of interest $t=T^{\mathrm{pred}}$ (e.g., 2 hours from now). Formally, the margin at time $t$ is: 
\begin{align}
\begin{split}
\min\{ 
      r^{\mathrm{up}}_t-\sum_i ({x}_{it} - \hat{x}_{it}),~
      r^{\mathrm{dn}}_t-\sum_i (\hat{x}_{it} - x_{it})\}.
\end{split}\label{eq:margin}
\end{align}
\end{subequations}
In \eqref{eq:margin}, the generic notation $x_{it}$ is used to represent all flexibile resources such as generators ($g^{\mathrm{p}}_{it}$), storage ($p^{\mathrm{ch/dis}}_{kt}$), load curtailment ($d^{\mathrm{curt}}_{nt}$) and the system margin is defined as the total sum of remaining flexibility in all units.

\subsection{Pre- and Post-processing}\label{subsec:processing}

To increase and validate the model performance, additional steps are added in the kernel SVR. First, the input data is normalized by each feature (dispatch signals and network status) so that they can be accounted for equally. Once the fitting of the kernel SVR is completed, then the performance is validated with the testing data.

\section{Numerical experiments}

\begin{figure}[t]
    \centering    
    \vspace{-0mm}
    \includegraphics[width=\columnwidth]{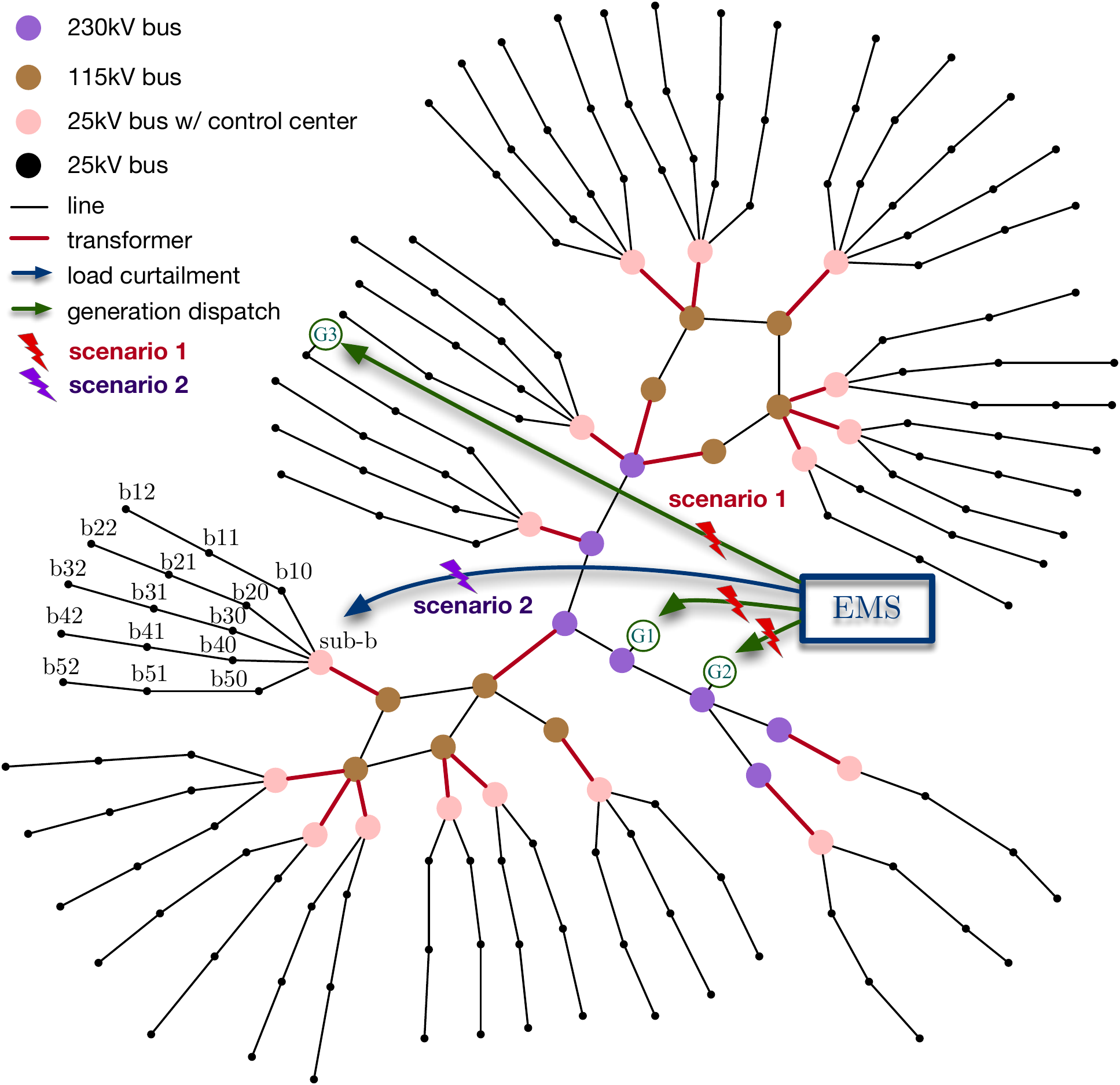}
    \vspace{-5mm}
    \caption{\small HCE 187-bus test system where nodal voltage levels are marked color-coded. Two intraday FDI attack scenarios \textendash~generation dispatch (scenario 1) and load curtailment (scenario 2) \textendash~in Numerical experiments are illustrated with navy color. \vspace{-0mm}}
    \label{fig:HCEnetwork}       
\end{figure}

We use the 187-bus test system shown in Fig.~\ref{fig:HCEnetwork} with daily generation and load profiles based on real-life data provided by Holy Cross Energy, a power utility in Colorado. The PV generation has been scaled up by a factor of six so that the overall solar penetration is around 15\% of total generation, which is similar to the current practice in the state of California (15.43\% of the total generation was from solar in 2020) \cite{CA_gen}. Demand forecast error is assumed to follow a Gaussian distribution with zero mean and 2\% of nominal value as standard deviation. For the attack model in \eqref{Eq:AttackerModel}, the generation dispatch (Scenario 1) and load curtailment (Scenario 2) are assumed to be falsified respectively. For the detection model in \eqref{Eq:SVR_original}, the monitoring window (${\cal T}^{\mathrm{m}}$) and the time of interest (${T}^{\mathrm{pred}}$) are set as six hours and two hours from the time of prediction, i.e., ${\cal T}^{\mathrm{m}}\!=\![T^{\mathrm{pred}}\!-8\text{h},~T^{\mathrm{pred}}\!-2\text{h}]$. The RBF kernel is used for feature map $\phi(\cdot)$. All optimization problems are modeled using Julia/JuMP  \cite{DunningHuchetteLubin2017}, and solved by Ipopt Solver \cite{wachter2006implementation}. The kernel SVR is implemented using the scikit-learn package \cite{scikit-learn}.

\begin{figure}[p!]
\begin{subfigure}[t]{\columnwidth}
    \centering    
    \vspace{-0mm}
    \includegraphics[width=\columnwidth]{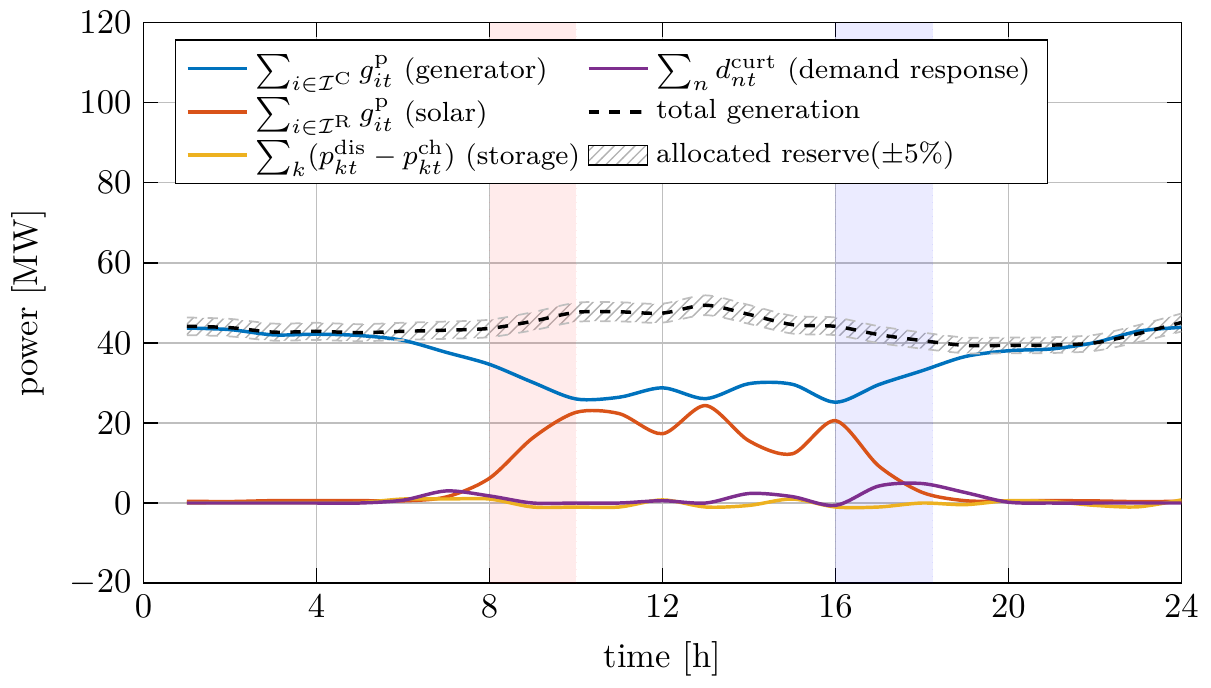}
    \vspace{-7.5mm}
    \caption{\small Illustration of generation dispatch and allocated reserve (5\% of the system load) on a \textit{cloudy day}. The red and blue boxes represent the net-load morning ramp down and evening ramp up periods. \vspace{-0mm}}
\end{subfigure}
 \vfill
 \begin{subfigure}[t]{\columnwidth}
    \centering    
    \vspace{-0mm}
    \includegraphics[width=\columnwidth]{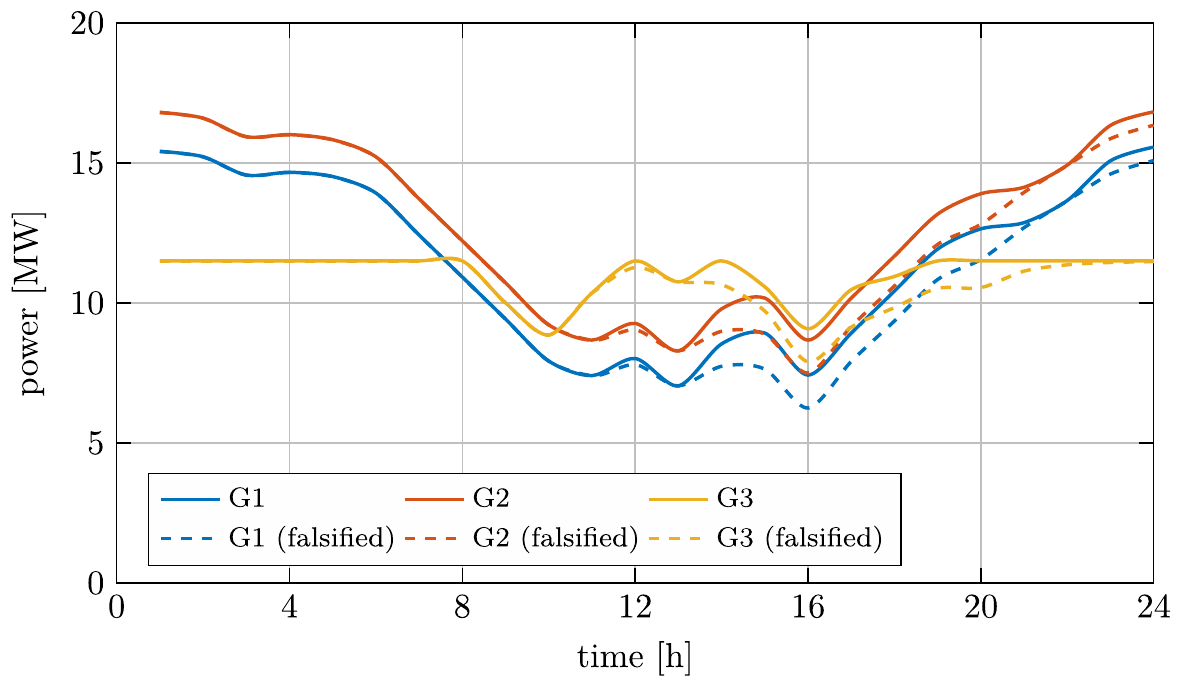}
    \vspace{-7.5mm}
    \caption{\small Comparison between the original scheduled generation dispatch (solid lines) and falsified FDI attack (dashed lines).  \vspace{-0mm}}
\end{subfigure}
 \vfill
 \begin{subfigure}[t]{\columnwidth}
    \centering    
    \vspace{-0mm}
    \includegraphics[width=\columnwidth]{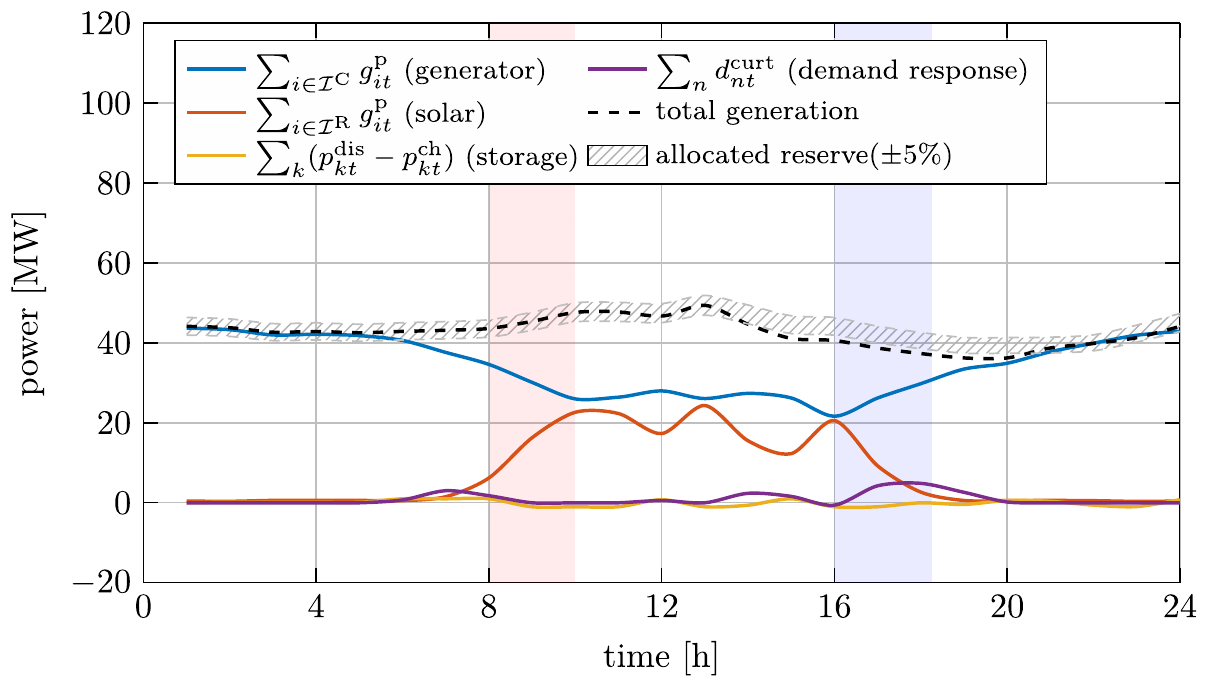}
    \vspace{-7.5mm}
    \caption{\small Generation profile and allocated reserve (5\% of the system load) under \textit{the intraday FDI attack on generation dispatch}.  \vspace{-0mm}}
\end{subfigure}
 \vfill
 \begin{subfigure}[t]{\columnwidth}
    \centering    
    \vspace{1.5mm}
    \includegraphics[width=\columnwidth]{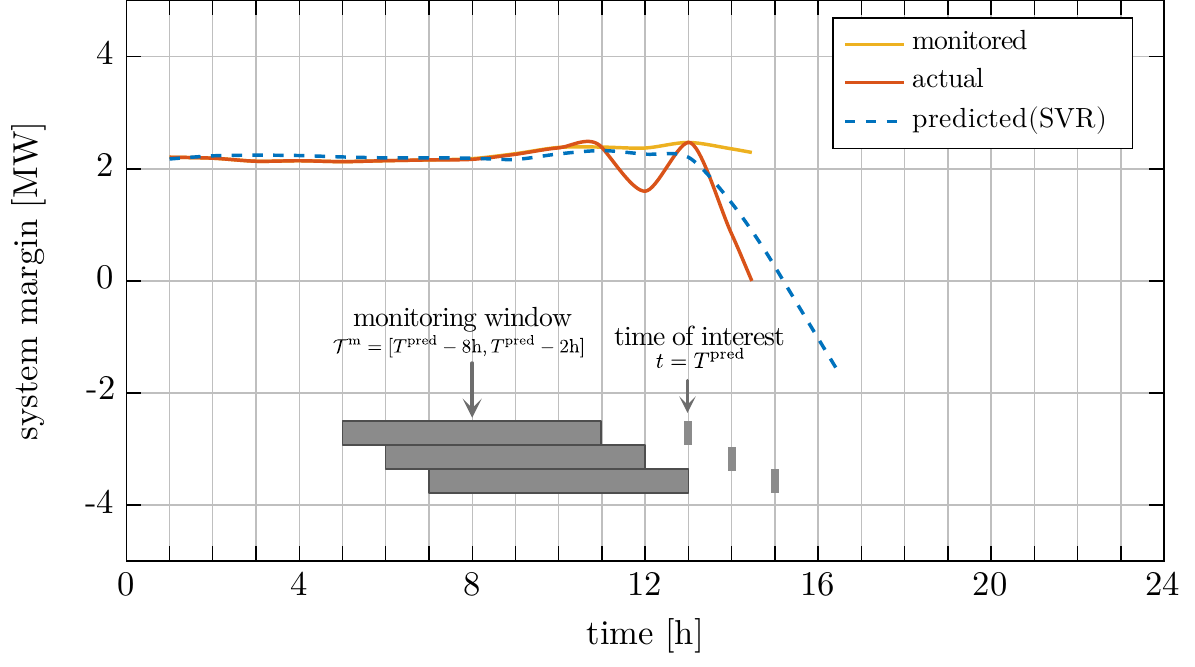}
    \vspace{-7mm}
    \caption{\small System margin value comparison under \textit{the intraday FDI attack on generation dispatch.}: values on EMS monitor, actual margin and predicted values from the detection model. \vspace{-1mm}}
\end{subfigure}
\caption{\small Illustration of Scenario 1: (a) dispatch prediction, (b) generation dispatch falsification, (c) generation profile under the intraday FDI attack, (d) system margin prediction.\vspace{-1mm}}
\label{fig:case1}
\end{figure}

\subsection{Scenario 1: intraday FDI attack on generation dispatch}\label{subsec:case1}

\subsubsection{Dispatch prediction}\label{subsubsec:case1_dispatch}
Given solar generation and demand forecasts from a cloudy day, the attacker solves the prediction model in \eqref{Eq:AttackerPred} and the resulting generation profile is shown in Fig.~\ref{fig:case1}(a). The solar generation (red line) has two valleys around 12:00 and 15:00 (due to the weather conditions such as intermittent clouds), which are offset mainly by the generators (blue line). In order to compensate for the change in solar generation output around sunrise and sunset, controllable resources such as gas generation, energy storage, and load curtailment are utilized. Additionally, the system reserve is allocated to address the system uncertainty.

\subsubsection{Falsification of generation dispatch}\label{subsubsec:case1_fal}
The attacker generates falsification signals by solving \eqref{Eq:AttackerModel} with the predicted dispatch. Fig.~\ref{fig:case1}(b)\textendash(c) show a scenario of the intraday FDI attacks on generation dispatch $g^{\mathrm{p}}_{it}$. In Fig.~\ref{fig:case1}(b), the original dispatch is planned to provide approximately 10MW/3hr of ramping-up flexibility around  sunset (16:00\textendash19:00). The attack signal overrides the original values (solid line) and instead injects reduced generation dispatch (dashed line). As a result, compared to Fig.~\ref{fig:case1}(a), the total system generation (dashed black line) in Fig.~\ref{fig:case1}(c) decreases due to this FDI attack, and the total amount of supply-demand imbalance exceeds the system flexibility (shaded area) which can cause a system-wide power outage.

\subsubsection{Detecting the FDI attack with kernel SVR}\label{subsubsec:case1_detection}
Fig.~\ref{fig:case1}(d) shows how the system margin changes over time when the FDI attack presented in Fig.~\ref{fig:case1}(b) is carried out. While the actual system margin (orange line) drops as a result of the attack and is eventually exhausted at 14:30, the EMS monitor (yellow line) shows a steady margin throughout the entire attack. This is because the attacker injects the falsified signal into the up-link to the EMS monitor as well as the down-link to the DERs. 
The proposed detection model provides the estimation on the system margin two hours ahead of the time of interest. For example, in Fig.~\ref{fig:case1}(d), the system margin prediction (blue dashed line) plotted for 13:00 is made at 11:00 based on the monitoring window 6:00\textendash12:00 and the prediction gets updated and shifted as time flows (the three gray blocks in the bottom of Fig.~\ref{fig:case1}(d) illustrate how the monitoring window and time of interests change over time). Thus, the grid operator can monitor this prediction and take preventive measures in a timely manner once the margin drops below the reliability threshold, i.e., at 13:00, the operator can know that margin is expected to drop below 1MW in an hour. Note that the monitoring and actual margin values are not recorded after the power outage at 14:30 and therefore the prediction is also available only until 16:30.

\begin{figure}[p!]
\begin{subfigure}[t]{\columnwidth}
    \centering    
    \vspace{-0mm}
    \includegraphics[width=\columnwidth]{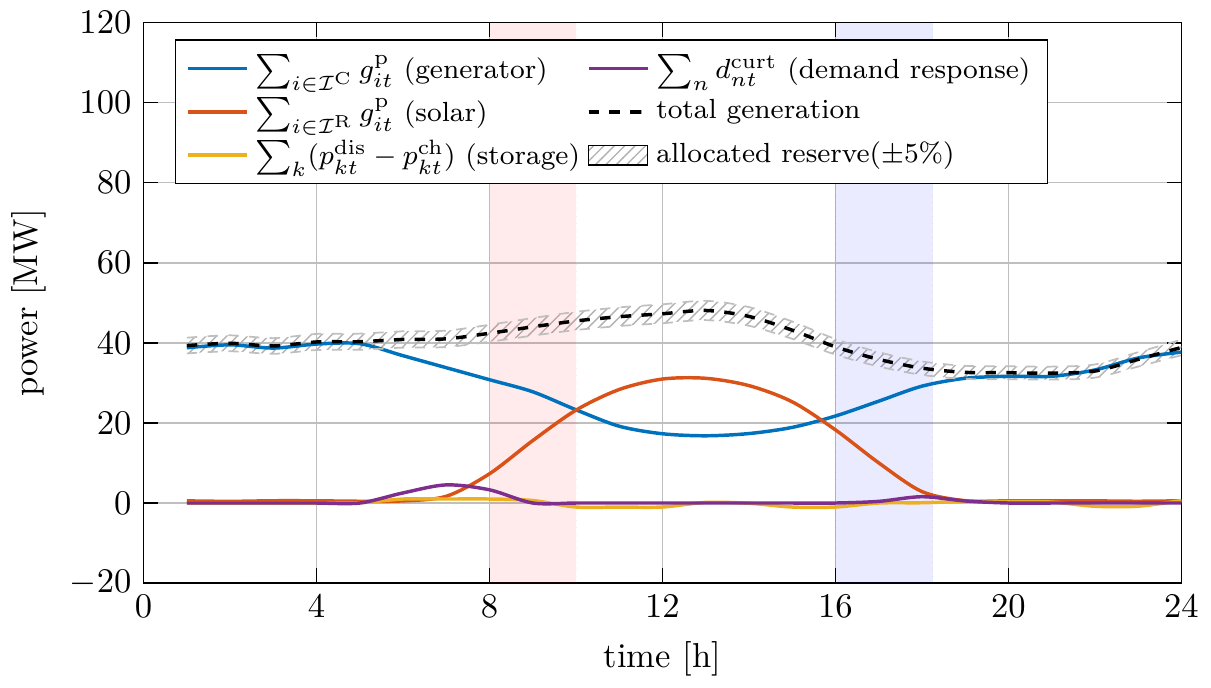}
    \vspace{-7mm}
    \caption{\small Illustration of generation dispatch and allocated reserve (5\% of the system load) on a \textit{sunny day}. 
    The red and blue boxes represent the net-load morning ramp down and evening ramp up periods.
    \vspace{-0mm}}
    \label{fig:GenProfile_normal} 
\end{subfigure}
 \vfill
 \begin{subfigure}[t]{\columnwidth}   
    \centering    
    \vspace{3.5mm}
    \includegraphics[width=\columnwidth]{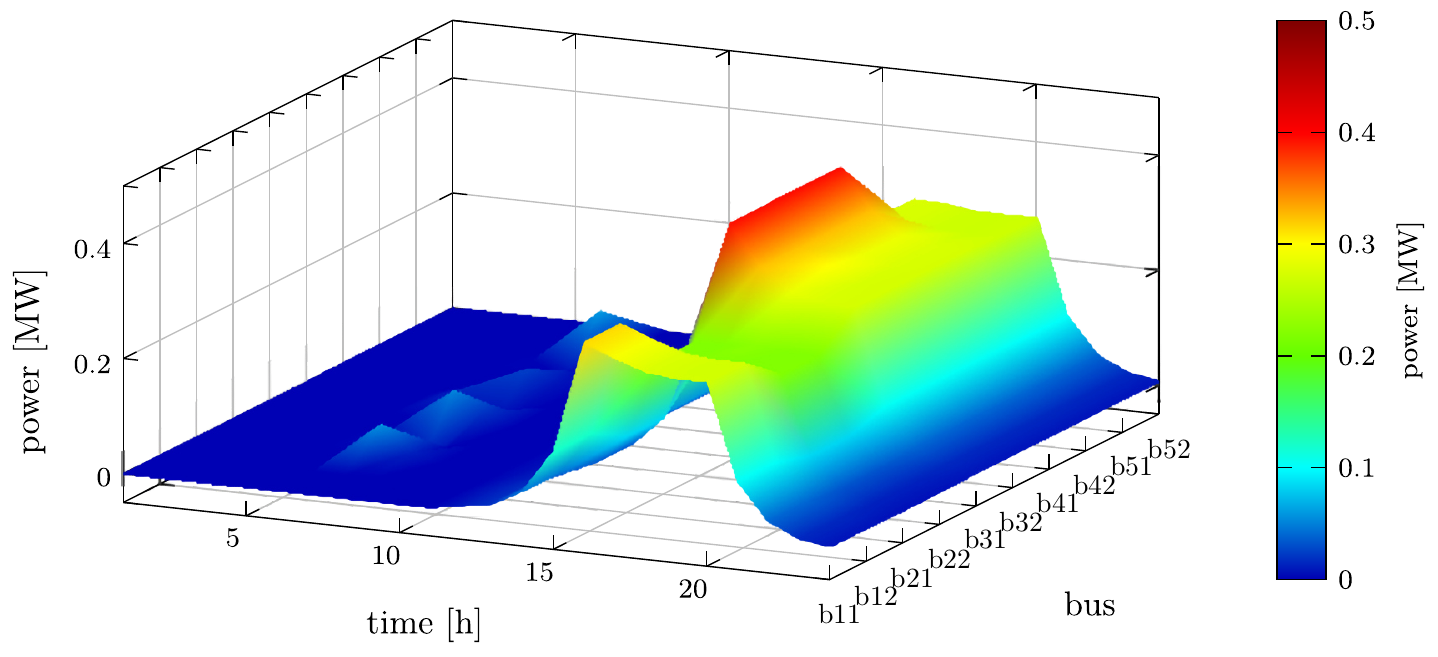}
    \vspace{-7.5mm}
    \caption{\small Falsified load curtailment signals on target load buses under \textit{sub-b} substation.   \vspace{-0mm}}
    \label{fig:FDIAsignal_dcurt}       
\end{subfigure}
 \vfill
 \begin{subfigure}[t]{\columnwidth}
    \centering    
    \vspace{3.5mm}
    \vspace{1mm}
    \includegraphics[width=\columnwidth]{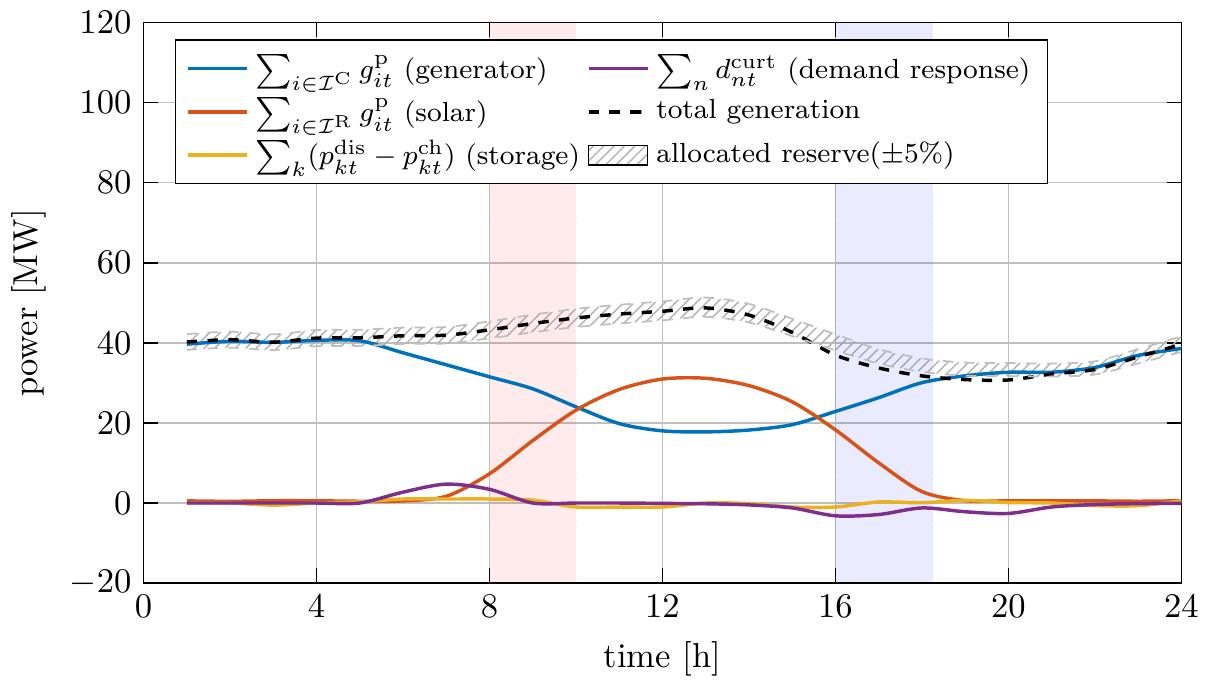}
    \vspace{-7.5mm}
    \caption{\small Daily generation profile and allocated reserve (5\% of the system load) under \textit{the intraday FDI attack on load curtailments.} \vspace{-0mm}}
    \label{fig:GenProfile_falsified_dcurt}  
\end{subfigure}
 \vfill
 \begin{subfigure}[t]{\columnwidth}
    \centering    
    \vspace{3.5mm}
    \includegraphics[width=\columnwidth]{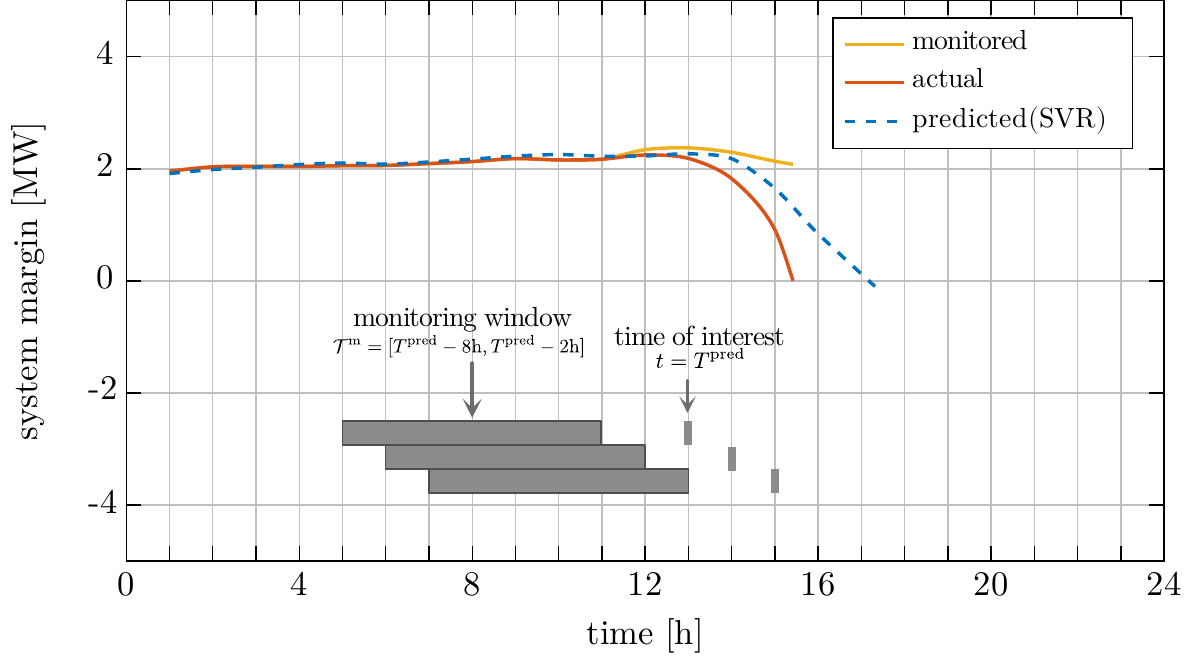}
    \vspace{-7.5mm}
    \caption{\small System margin value comparison under \textit{the intraday FDI attack on load curtailments.}: values on EMS monitor, actual margin and predicted values from the detection model. \vspace{-1mm}}
    \label{fig:margin_dcurt}  
\end{subfigure}
\caption{\small Illustration of Scenario 2: (a) dispatch prediction, (b) load curtailment falsification, (c) generation profile under the intraday FDI attack, (d) system margin prediction.\vspace{-1mm}}
\label{fig:case2}
\end{figure}

\subsection{Scenario 2: intraday FDI attack on load curtailments}

\subsubsection{Dispatch prediction}\label{subsubsec:case2_dispatch}
Fig.~\ref{fig:case2}(a) shows the generation dispatch and demand profile of a sunny day predicted by an attacker. Unlike Scenario 1, the solar generation shows a smooth ramp-up during the sunrise and ramp-down during the sunset, and in turn, the generation output covering the variability from solar is also expected to have a smooth shape without a valley.

\subsubsection{Falsification of load curtailment}\label{subsubsec:case2_fal}
In this scenario, we assume the control center at \emph{sub-b} in Fig.~\ref{fig:HCEnetwork} is hacked and the downstream load curtailment signals ($d^{\mathrm{curt}}_{nt}$) are falsified. In contrast to the falsification of generation dispatch signals in Fig.~\ref{fig:case1}(b), the FDI attack on load curtailment requires manipulation on a greater number of signals that are geographically related (nearby electricity loads have similar patterns). To make the falsification signal more natural, the attacker adds a geographical regularization in addition to the temporal regularization term in the objective function and solves the attack model in \eqref{Eq:AttackerModel} to design an attack. Thus, the falsification signals ($\Delta d^{\mathrm{curt}}_{nt}$) on nearby nodes (e.g., \emph{b11} and \emph{b12}, \emph{b51} and \emph{b52}) in Fig.~\ref{fig:case2}(b) have a similar shape over time, which makes the attack hard to be detected. As a result of the attack, as shown in Fig.~\ref{fig:case2}(c), the total system generation (dashed-line) decreases gradually and the total amount of supply-demand imbalance exceeds the system flexibility (shaded area).

\subsubsection{Detecting the FDI attack with the kernel SVR}\label{subsubsec:case2_detection}
Similar to Scenario 1, the monitored system margin (yellow line) in Fig.~\ref{fig:case2}(d) looks normal throughout all intervals. However, the actual margin (red line) starts to decrease at 13:00 and is fully exhausted at 15:20. The proposed kernel SVR detection model predicts the drop would begin at 14:00 and the margin will be below 1MW at 15:30. In other words, the grid operator will notice the change at 12:00 (two hours ahead of 14:00) and the preventive measure will be taken at 13:30 (two hours ahead of 15:30) if the security threshold is 1MW.

\section{Conclusions and Future work}\label{sec:conclusion}
This paper analyzed the vulnerability of power grids with high PV penetration against an intraday FDI attack that falsifies DER dispatch and monitoring signals. Based upon the dispatch prediction and dispatch falsification models, we illustrated how gradual manipulation of DER outputs can cause a power imbalance which exceeds the system reliability margin. To enhance the power grid reliability against the attack scenario, we also proposed a detection model utilizing a kernel SVR which allows a power grid operator to predict the reduction in the system margin ahead of time. The numerical experiments demonstrate the attack scenarios and the performance of the detection model on the HCE test system, which is based on real-world demand and generation profile data provided from a power utility in Colorado.

There are several directions for future works. First, we plan to consider the relaxation of perfect knowledge assumption on the grid conditions (nodal voltage and phase angle). In practice, such information is available only in limited locations and we will investigate how this affects the detection model. Second, we plan to evaluate the performance of the kernel SVR detection model against other detection methods such as state estimation. Lastly, we plan to implement the intraday attack and the kernel SVR detection model on the demonstrator for real-life verification.

\section*{Acknowledgement}
The authors would like to thank Bruno Leao and Ulrich Muenz at Siemens Technology for helpful discussions regarding the attack scenarios. We thank Chris Bilby at Holy Cross Energy for sharing relevant datasets.

\balance
\bibliographystyle{IEEEtran}
\bibliography{fdia_cdc.bib}
\end{document}